\documentclass[amsmath,amssymb,reprint]{revtex4-1}
\usepackage{graphicx}
\usepackage{dcolumn}
\usepackage{bm}
\usepackage[utf8]{inputenc}
\usepackage{float}
\usepackage{subfigure}
\usepackage{xcolor}
\usepackage[export]{adjustbox}[2011/08/13]
\usepackage{esvect}

\makeatletter

\renewcommand*{\p@subsection}{}

\renewcommand*{\p@subsubsection}{}
\makeatother

\begin{document}
	
\title[Cost-Based Approach to Complexity: A Common Denominator?]{Cost-Based Approach to Complexity: A Common Denominator?}
	\author{L. da F. Costa}
	\affiliation{S\~ao Carlos Institute of Physics, University of S\~ao Paulo, S\~ao Carlos, SP, Brazil}
	
	\author{G. S. Domingues}
	\affiliation{S\~ao Carlos Institute of Physics, University of S\~ao Paulo, S\~ao Carlos, SP, Brazil}
	
	\date{\today}
	
	\begin{abstract}
		Complexity remains one of the central challenges in science and technology. Although several approaches at defining and/or quantifying complexity have been proposed, at some point each of them seems to run into intrinsic limitations or mutual disagreement. Two are the main objectives of the present work: (i) to review some of the main approaches to complexity; and (ii) to suggest a cost-based approach that, to a great extent, can be understood as an integration of the several facets of complexity while keeping its meaning for humans in mind. More specifically, it is poised that complexity, an inherently relative and subjective concept, can be summarized as the cost of developing a model, plus the cost of its respective operation.  As a consequence, complexity can vary along time and space.  The proposal is illustrated respectively to several applications examples, including a real-data base situation.
		
	\end{abstract}
	
	\maketitle
	
	\section{Introduction}
	
\begin{figure}[!btp]
		\centering
		\includegraphics[width=0.9\linewidth]{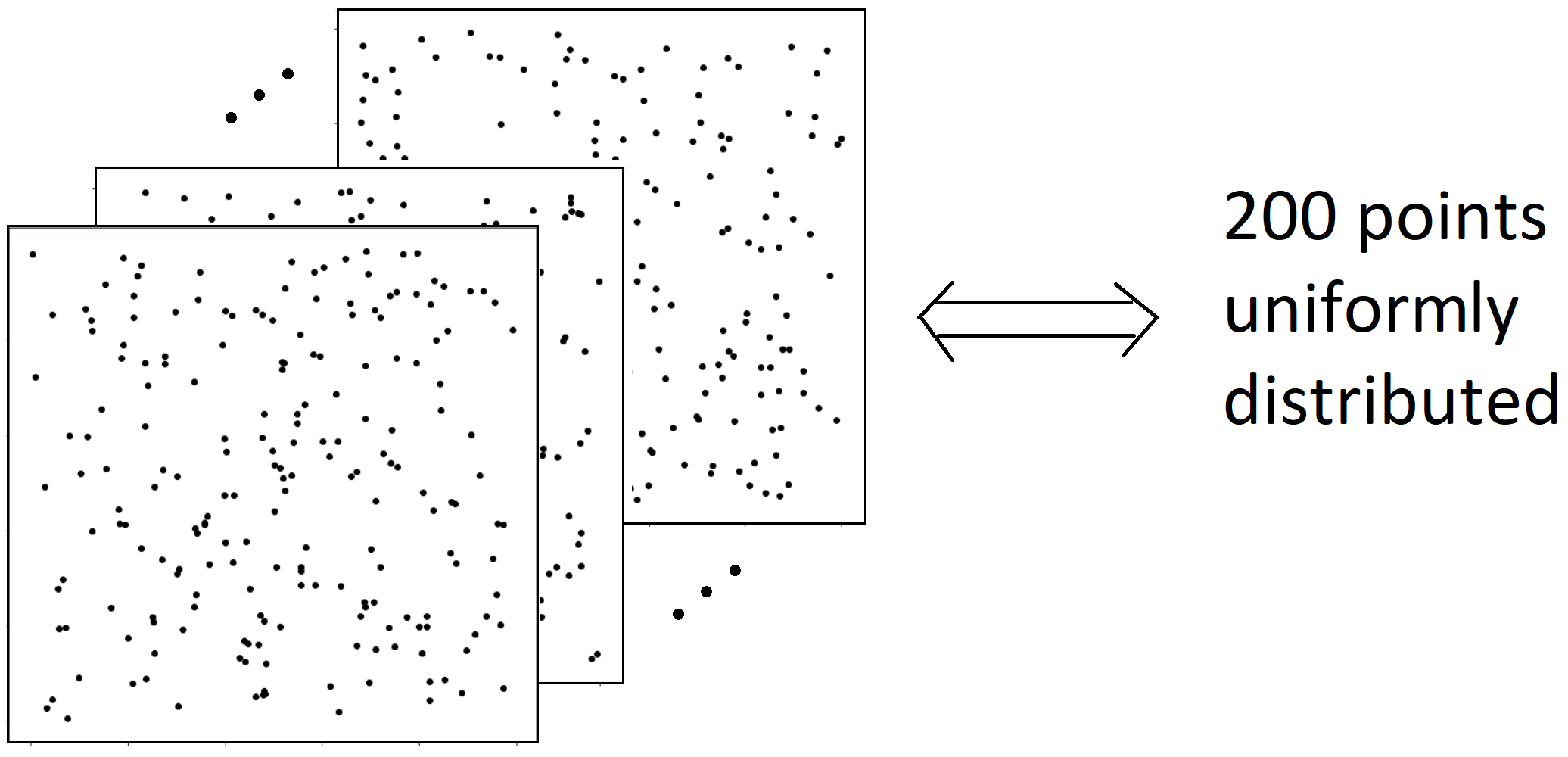}
		\caption{The size of an entity is one of the most intuitive attempts at measuring complexity. However, this concept may run into difficulties, such as in the case of a distribution of many points on a space uniformly distributed. Despite its large size, this distribution is by every means very simple, being describable by a simple sentence. Observe that the quantification of complexity often involves the mapping of an entity from one space (typically nature) into another (e.g. language, logic and/or mathematics).  In addition, the complexity may also have to take into account eh
			computational expenses required for implementing each of the mentioned possibilities.}
		\label{UniformDistribution}
	\end{figure}

	One of the most often mentioned terms in science currently is \emph{complexity}. Though we have an intuitive understanding of this concept, to the point of often being able to readily recognize if something is complex or not, it turns out that it is particularly difficult to objectively define complexity (e.g. \cite{Heylighen1996},~\cite{Edmonds1995},~\cite{Ferreira2001}). Indeed, several of the approaches that have been proposed for defining and better understanding complexity sooner or later run into intrinsic limitations. For instance, we can attempt to define the complexity of a given set of 10 images containing 200 points each distributed uniformly, as illustrated in Figure~\ref{UniformDistribution}.  This data can be characterized in at least three manners: (i) by storing all the complete images, requiring 1280 bytes; (ii) by obtaining the position of each of the points in each of the images in the set, requiring approximately 15.6 Kb, (iii) by just indicating that the whole set corresponds to 10 realizations of an uniform random process with 200 points.
	It should be kept in mind that, though we limited our discussion to just three
	possibilities, there is a virtually infinite number of alternatives, which contributes to making complexity even more complex.
	
	In the above examples, in which complexity is quantified in terms of the respectively required memory, we reach three distinct conclusions, corresponding to large, small, and medium complexities.  Though it would be tempting to conclude that the complexity of the problem is ultimately that of the shortest of the three descriptions, there are additional intricacies to be considered.  For instance, choices (i) and (ii) will both allow the recovery of the original data, which cannot be achieved from the description obtained by option (iii).  So, in case the positions of the points do matter for a specific application, we need to conclude that the original data is not so simple and perhaps assign the complexity corresponding to the memory size implied by option (ii).  However, if the position of the points is irrelevant for a given problem, we can conclude that the original data is indeed extremely simple.  A number of additional constrains can influence on the above analysis.  For instance, it may happen that we are also concerned with the computational cost required for achieving the different levels of compaction, and so on.

	The frequent conceptualization of complexity in terms of the length of the respective description reveals a close relationship between complexity and scientific modeling, more
	specifically the level of \emph{abstraction}. As a matter of fact, accurately describing a phenomenon as an abstract construct consists in one of the main objectives of modeling. Another important one is to provide subsidies for making predictions about the observed phenomenon. Such modelings imply mapping the real world into a relatively precise and formal system of representations, such as natural language and/or logic and/or mathematics. 
	
	In addition to the above discussed intricacies in defining complexity, it is also important to observe that another critical aspect that has severely constrained attempts at defining complexity concerns the fact that complexity is ultimately a human-related concept, therefore involving some level of subjectivity required for more generality.  Indeed, any more comprehensive definition of complexity has to remain compatible with the intuitive understanding of this term by humans in general.  The hypothesis here is that, though humans may have specific differences in their understanding of complexity at large, there should be some common elements shared by a substantial number of people.  The alternative of proposing a definition and characterization of complexity purely from mathematical and physical considerations has proven to be rather difficult to be accommodated generally into the human common sense, as our simple discussion above has already illustrated.  These mathematical approaches are nevertheless welcomed and fine in themselves, as they provide additional insights into more specific aspects of complexity.  However, here we aim at developing a more comprehensive approach that would be largely coherent with the understanding of complexity shared by humans.
	
	The approach to complexity developed in this work is aimed specifically at trying to summarize as best, objective and quantitatively as possible its human connotation.  This has been achieved at the expense of obtaining a strict, fully accurate and objective mathematical definition.  Instead, in the present approach we resource to the \emph{costs}, \emph{expenses} or even the \emph{difficulty} of developing and operating models of the real-world.  The intrinsic advantage of this approach is that it paves the way for accommodating the fact that complexity, as understood by humans, often varies substantially along time and space, depending on the available resources and specific constraints.  Indeed, the very concept of cost was very probably developed precisely to account for providing a quantification of these varying demands and constraints.  What is complex today, will probably be simpler tomorrow, and what is complex for a given scientific team in a given place may be simpler for a team in another place that is more acquainted with the respective area of a problem, and/or having access to more resources.  
	
	In brief, the suggested approach to complexity aims at achieving as much compatibility with its human connotation at the expense of a more strict mathematical definition which would probably incur in being too specific and not able to accommodate the fact that complexity as understood by humans can vary along time and space.
	
	We start by providing a brief review of some of the various previous attempts that have been made to define complexity, and then present our proposal of a potentially new way of looking at and understanding complexity in terms of model and operation costs, as well as how the efficiency of these solutions can be related to complexity. Four case examples are then discussed in order to better illustrate the proposed approach.

	\section{Some approaches to complexity}
	
	One important issue to be considered from the outset is that there are two aspects to complexity: (a) definition; and (b) quantification.  In particular, the former seems to be to a large extended implied by the latter because if one is capable of measuring complexity, we can promptly decide on a phenomenon being complex or not. Yet, some of the definitions of complexity are predominantly qualitative. 
	
	In this section, we provide a concise review of some of the main approaches that have been advanced for quantifying (and therefore defining) complexity. It should be observed that this review is not  fully comprehensive, and additional interesting approaches can be found.

	\textbf{Informational Complexity:} Information Theory~\cite{shannon1948} studies the usage and transformation of information, as well as its transmission. Information is often approached in terms of messages involving sets of symbols. Deriving from thermodynamics and information theory, the concept of \emph{entropy} allows an effective statistical means for quantifying the amount of information (e.g. in bits) of a set of symbols. Let's consider the Shannon entropy,~\cite{shannon1948} typically measured in \emph{bits}, given as
	\begin{equation}
		E = -\sum_{i = 1}^{S} p_{i}\log_{2}(p_{i})
	\end{equation}
	
	where S is the number of involved symbols and $p_i$ their respective probabilities or relative frequencies. For instance, if we have a text containing 50 times the word `tea' and 50 times the word `time' (observe that $S = 2$), we will need, in the average, 1 bit for representing the information in this set. However, if we change the number of instances to $10$ and $90$, respectively, we have an average minimum of only approximately $0.469$ bits. 
	
	It can be verified that the situation when all symbols have the same probability leads to maximum entropy, while its minimum value is observed when only a symbol has non-zero probability. The use of the average minimum of bits obtained from entropy provides an interesting approach to quantify the complexity of an entity represented as a set of symbols, and can often lead to satisfactory results. Examples of usage can be found in~\cite{carothers2004,lin1991,wu2011}. However, this approach typically does not consider the interrelationship between the involved symbols (other types of entropy can be used here) and, more importantly, a sequence of $S$ symbols drawn with uniform probability will yield maximum entropy, while being statistically trivial (such sets can be obtained by sampling the uniform distribution, one of the simplest density of probability).  In other words, maximum information (and complexity) can be easily obtained from relatively simple generative models.
	
	\textbf{Geometrical Complexity:} Perhaps as a consequence of being more directly perceived, the complexity of visual patterns, shapes, and distribution of points/shapes, has attracted great attention from the scientific community. While a dot and a straight line can be conceptualized as exhibiting minimal complexity, structures such as the border of islands, snowflakes and some types of leaves are characterized by intricate self-affine geometries.  Thus a notion of complexity can be developed by estimating how much these relatively more sophisticated shapes depart from  simpler ones (dots, lines, filled regions, etc.). Several approaches have been proposed for characterizing geometrical complexity, as in porous media analyses,~\cite{xia2019} especially the concepts of fractal dimension (e.g. \cite{peitgen2006chaos,niemeyer1984}) and lacunarity (e.g. \cite{pandini2005self,dong2000}). Briefly speaking, fractal objects exhibit self-affine structure extending over all (or a wide range of) spatial scales, therefore imparting high levels of complexity to such objects. Observe that the fractal dimension takes real values not necessarily corresponding to a topological dimension (limited to integer values). 
	
	The concept of lacunarity, which was proposed by B. Mandelbrot in order to complement the fractal characterization of objects, expresses the degree of translational (or positional) variance of an object while observed at varying spatial scales (e.g. \cite{pandini2005self}), being used to quantify the capability of the object to occupy empty spaces inside its own geometry.  The lacunarity can also be understand as "gappiness", indicating how much a figure's texture has "holes" or is inomogenous.~\cite{allain1991,kaye2008}
	
	Another important measurement in terms of geometrical complexity is the succolarity, which measure the connectivity of pores or gaps in an image or object and, alongside with lacunarity, has been extensively used to differentiate objects with the same fractal dimension.~\cite{xia2019,anovitz2015}
		
	The fractal/lacunarity approaches can provide valuable information in many situations and with respect to a wide range of data.  However, similarly to entropy, it is also possible to identify simple generative rules (e.g.~Koch curves or Mandelbrot sets) that will yield self-affine structures with relatively large fractal values.
	
	\textbf{Computational Complexity:} One of the interesting approaches that have been proposed to define and characterize complexity involves the concept of computational complexity (e.g. \cite{Papadimitriou1993}). 
	
	Given a specific computation, the respective order of complexity quantifies the amount of computational resources (typically processing time and/or memory capacity) required for its effective calculation. For example, adding two vectors containing $N$ elements each is characterized by computational complexity order of $O(N)$, where $O()$ stands for the `big O' notation. In this particular example, it is meant that adding the two vectors will involve a number of additions proportional to $N$. Observe that there are some intricacies in determining the $O()$. For instance, adding three vectors with $N$ elements each will imply $2N$ additions, but we still get the same $O(N)$ for this case. It is not often easy to calculate the $O()$ of a given operation, and the reader is referred to the respective literature for more information on this important and interesting area (e.g. \cite{Papadimitriou1993,cooper1990,monasson1999}). While the order of complexity provides a formal way to quantify some aspect of the complexity of a computation, it cannot be directly applied to characterizing the complexity of entities and it may not be known or determinable in certain situations.
	
	Though computational complexity provides substantially important subsidies for classifying types of algorithms, it is possible to think of simple programs which have high computational complexity.  For instance, the concept and code for calculating the $P$ power of an $N \times N$ matrix is short and simple, but involves a relatively high computational complexity of $O(PN^2)$.  It should be observed that this operation can be performed more effectively after the matrix is diagonalized, implying in a substantially more complex code.
	
\begin{figure}[!htbp]
		\centering
		\includegraphics[width=0.7\linewidth]{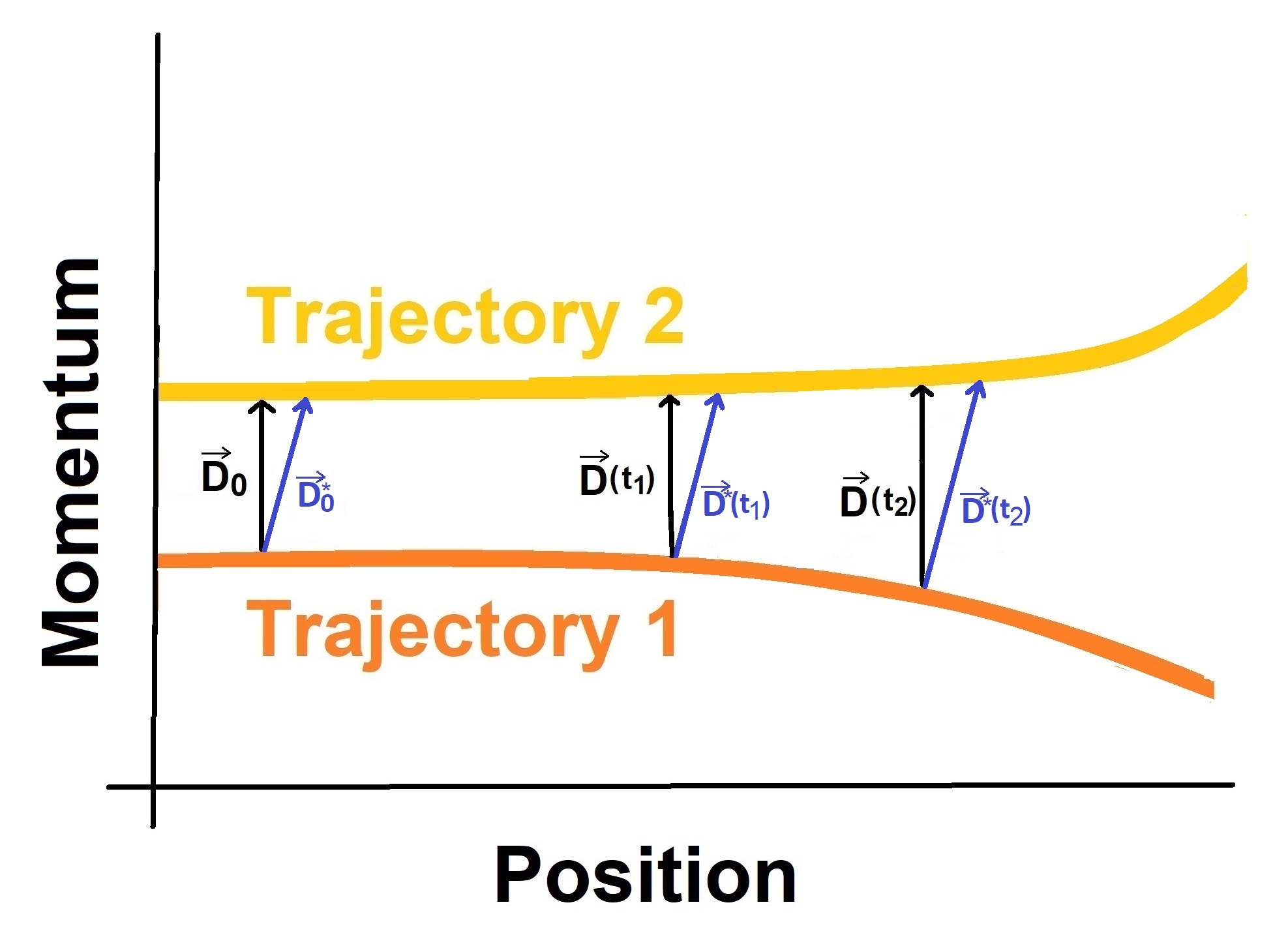}
		\caption{A \emph{phase space} with two infinitesimally close trajectories departing one from the other. Here we illustrate two examples of how the separation vectors $\vec{D}(t)$ and $\vec{D}^*(t)$ can be measured.}
		\label{PhaseSpace}
	\end{figure}

	\textbf{Transients and Steady-states:} The state of a given system can be characterized by a set of constants and measurements or dynamic variables of that system. We can also use the values of these variables to define a point in the \emph{phase space} corresponding to that state (see Figure~\ref{PhaseSpace}).  The phase space can be understood as a space that corresponds to the possible states of the dynamic variables involved in a dynamical system (e.g.~\cite{nolte2010}).  The dynamics in a system can be at two main types of regimes: \emph{transient} or \emph{steady-state}. 
	
	The \emph{Steady-state} can also be characterized by a configuration of parameters that makes the state of a given system invariant in time, or locally restrict in the phase-space (e.g. an oscillation), as it reaches an equilibrium.~\cite{moxnes2016}
	
	However, throughout the evolution of a system, it may have a period along which its parameter and dynamical variables suffer alterations in order to allow the system to a change in the phase or state, going from a regime to another. In the phase-space, that period is usually called a \emph{transient}.~\cite{moxnes2016}
	
	\textbf{Dynamical Systems Complexity:} The area of dynamical systems has been extensively (e.g. \cite{peitgen2006chaos,waldrop1993complexity,kauffman1996home,d2021complexity,rosenkrantz2021,kinouchi2020}) developed in order to represent the interaction along time between the components of a given system. One of the main concerns in that area is to identify if the systems dynamics will converge to a steady state in the long term and how the long term behavior then depends on its initial condition.~\cite{rickles2007}
	
	Another related concept is the Lyapunov's coefficient,~\cite{boeing2016,angelo2009} which measures how fast the distance between two infinitesimally close trajectories in the phase space increases when the trajectories depart one from the other.  This coefficient, henceforth expressed as $\lambda$, can be defined in terms of the following equation:
	
	\begin{equation}
		|\vv{D}(t)| \approx e^{\lambda t}|\vv{D}_0|
	\end{equation}
	
	where $\vv{D}(t)$ is an infinitesimal separation vector between two close trajectories at a given time $t$. However, the orientation of the initial separation vector $\vv{D}_0$ can influence the value of $\lambda$, motivating the alternative approach called \emph{spectrum of Lyapunov's coefficients}, i.e. a set of values obtained for the separation rate of two trajectories in terms of several possible orientations, equal in number to the dimensionality of the phase space. Usually the largest Lyapunov's coefficient is chosen to determine the level of predictability, or complexity, of a system. An example of numerical calculation of  Lyapunov's coefficients for the \emph{Lorentz model}~\cite{lorenz1963} can be found on~\cite{shimada1979}.
	
	Examples of application of this approach include population models (such as the logistic approach) and the behavior of oscillators such as a pendulum. Though linear dynamical systems are relatively simple, non-linear counterparts can exhibit surprising dynamic characteristics, such as the fact that small perturbations in the system input can induce large variations of the respective output, a phenomenon that is associated to chaotic behavior.
	
	An important concept in Dynamical System Theory is that of an attractor, which is a set of numerical values that the system tend to recur along its dynamics.~\cite{grebogi1987}  Non-linear systems can have rather complex attractors, such as fractals, so it makes sense to speak of the complexity of a dynamics in terms of the complexity of its respective attractor. The \emph{spectrum of Lyapunov's coefficients} can also be used to estimate the fractal dimension of a respective attractor~\cite{kaplan1979,frederickson1983} and provide an upper bound for the information contained on a studied system.~\cite{Renyi1959} However, maximum unpredictability and disorder do not, necessarily, means high complexity (we have already seen that numbers drawn with uniform probability are easy to understand and model from the statistical point of view). Nevertheless, the dynamical system approach to complexity is particularly enticing in which concerns the idea that complexity would take place somehow at the mid point between simple, predictable dynamics and the highly unpredictable chaotic states. So, complexity would be mostly found at the border of chaos (e.g.~\cite{waldrop1993complexity,kauffman1996home}).

	\textbf{Self-Organized Criticality (SOC):} There are Dynamical Systems with many spatial degrees of freedom that have critical points as attractors, i.e. those systems spontaneously evolve into unstable states of the \emph{phase space}.  These so-called  \emph{Self-Organized Critical Systems}~\cite{bak1987} maintain a scale-invariance characteristic as they evolves towards non-equilibrium states around phase transition points via feedback mechanisms,~\cite{buendia2020} that often produces avalanche-like behaviors where a small perturbation can cause a wide change in the system. A typical numeric example is the \emph{sand pile model}, which is built on a finite grid where each site has an associated value that corresponds to the height of the pile. This height increases as "grains of sand" are randomly added onto the pile, until the height exceeds a threshold and cause the site to collapse, moving sand to neighbors sites, increasing their respective heights.~\cite{bak1987} The SOC model can be applied to many different situations.~\cite{reia2014}
	
	Other studies have shown that a system can have more than one critical point as attractor in a non-continuous phase transition, evolving to the coexistence of two unstable states and having avalanche-like behavior not only around one attractor, but also between both attractors on a larger scale, producing a cyclical effect. This type of system has been called Self-Organized Bistability (SOB).~\cite{buendia2020}
	
	The average avalanche size $\Delta S$ can be seen as an indicator of complexity on a system, and when $\Delta S \rightarrow \infty$ the system is considered to be on a critical state.  These properties can be understood as an indication of structural and dynamical complexity of a given system.  SOC is considered to be one of the mechanisms by which complexity appears in nature~\cite{bak1995}, being often applied in fields such as geophysics, ecology, economics, sociology, biology, neurobiology and others.~\cite{smalley1985,bak1990,Bak1996,beggs2003} 
	
	\textbf{Minimum Description Size (Kolmogorov Complexity):} Another interesting approach at defining/quantifying complexity considers the size or length of the minimal description of an entity or the resources necessary to reproduce that entity.~\cite{kolmogorov1963,kolmogorov1998} More formally speaking, this measurement takes into account the coding of an operation into a Turing machine.~\cite{copeland2004} The latter is an abstract, universal type of computing engine in which symbols are stored in an infinite tape that can be scanned by a head capable of performing some basic operations, also involving some other components such as state registers. 
	
	The Turing machine is often considered because it represents an abstract universal model of computing, but the quantification of description complexity can also be approached by considering other, more generally known, programming languages, such as C or Python, and hardware architectures, such as parallel, pipeline, GPU, etc. Thus, given an entity, we need to find the shortest program that can reproduce it. The complexity of that entity could then be gauged in terms of the length of the respective code (e.g. number of instructions) or the number of bits necessary to reproduce that code as a character string. 
	
	Let's consider the case of our distribution of points used in our Introduction section. Here, it would be  easy to obtain an extremely short program that produces that distribution. Such a program, in Python, could be as follows:
	
	\begin{tt}
		import random
		
		X = [random.random() for i in range (200)]
		
		Y = [random.random() for i in range (200)]
	\end{tt}
	
	Other examples of usage can be found in~\cite{mayordomo2002,petrosian1995,allender1992}. Though representing an interesting approach to quantifying complexity, the minimum description length depends intrinsically on the sequential type of coding and execution implied by the Turing machine. There are, however, many different computational paradigms, such as recursive (e.g. LISP) and parallel/distributed, that could be considered instead of the abstract Turing machine, implying in potentially very different code lengths. Even non-electronic means, biological, quantum, or even natural languages could be considered, implying completely different programming and storage organizations. A same problem, when programmed in such different computing systems, would present varying minimum description sizes. An additional difficulty is that it is often a challenge to find the minimum code for any problem.~\cite{vitanyi2020}
	
	While Shannon information theory is primary concerned with the information contained in messages of communication area, approaches to complexity based on Kolmogorov's minimal length tend to consider generative aspects of a given set of data.~\cite{grunwald2004,kolmogorov1983}
	
	\textbf{Bennett's Logical Depth:} This method can be informally understood (e.g.~\cite{Edmonds1995}) as a combination of the computational complexity and minimum description length approaches. More specifically, it corresponds to the computational expenses required for performing the minimal code obtained for reproducing the entity or phenomenon of interest. As such, this method focuses on the computational efforts required to reproduce a phenomenon or entity. 
	
	It is important not to confound this approach with Kolmogorov's complexity, that takes into account the length of the code and not the computational complexity or the execution time. By defining \emph{depth} as an effort of code execution, we have that an object that requires a long time to be reproduced cannot be quickly obtained by joining faster generated ones.~\cite{bennett1995} In other words, simplistically speaking, the divide-and-conquer approach would not lead to velocity gain in these cases, which would otherwise be better understood under the idea that the whole is larger than the sum of its parts.
	
	Though intrinsically interesting and with good potential, being useful in several situations and problems (e.g.~\cite{antunes2006}), this approach inherits to some extent the intrinsic limitations of the two approaches which it incorporates. For instance, if we considered the complexity order of the simple program we derived for producing our distribution containing $N$ points, we would obtain $O(2N)$, suggesting a large complexity for that otherwise simple set.
	
	\textbf{Network Complexity:} With the impressive development of the area of Network Science(e.g.~\cite{ALBERT}), aimed at studying complex networks, the concept of complexity has also become associated to the structure of networks used to represent a given entity or phenomenon. One of the reasons for the importance of networks science is the capacity of a graph or network to represent virtually any discrete system. For instance, networks can be used to model not only entities (e.g. airport routes, communication networks, scientific publications, links between web pages, etc.~\cite{costa2011}), but also procedures, e.g.~in terms of semantic networks (e.g.~\cite{Ferreira2001}) as well as several types of dynamics. The complexity of a network is related to how much its topology departs from that of regular or uniformly random networks (e.g.~\cite{Costa2018}). Generally speaking, a complex network tends to exhibit a non-trivial topology of interconnections. Such heterogeneities have to do not only with the node degree distribution, but also with many other topological features of the studied networks.~\cite{Costa2018}
	
	\textbf{Interpretation and Descriptive Complexity:} Löofgren~\cite{lofgren1973formalization}~\cite{lofgren1977complexity} describes an interesting approach to complexity involving the mapping from the system of interest into its respective description through learning, while the inverse mapping is often understood as interpretation.~\cite{Edmonds1995} This concept is combined with computational and description complexity, and a basic language is adopted to model the proposed framework. An alternative language-based approach to complexity has also been proposed in~\cite{Edmonds1995}.
	
	\textbf{Combinatorial Systems:} These are systems composed of components that can be redistributed to form new instances, or expressions, of that system following a set of rules, or \emph{grammar}, that determine which arrangements of components are allowed.~\cite{changizi2001} Common examples of combinatorial system are the human language, where a combination of a given number of words on a given order form a sentence; biochemistry,~\cite{dumartin2020} where different chemical components are combined to form new pharmacological compounds; or even institutions, like universities, which are formed by a combination of different departments. Given a number of possible expressions $E$ of a system, a number of different component types $C$ and an average expressions length $L$, these three characteristic parameters of a combinatorial system are usually related by the following relationship:
	
	\begin{equation}
		E\propto C^L
	\end{equation}
	Therefore, these systems can be classified accordingly to the variations observed for each parameter, that is, if $L$ is usually invariant while $C$ dictates the behavior of $E$, or if both $C$ and $L$ vary and equally impact the behavior of $E$, to describe just some examples.
	
	In the case of human language, for instance, as the number of words ($C$) and the length of the phrases ($L$) increases, therefore increasing the number of possible phrases ($E$), it can be expected that the complexity of that language will also increase, therefore being somehow proportional to $E$.
	
\begin{figure*}[!bt]
		\centering
		\includegraphics[width=\linewidth]{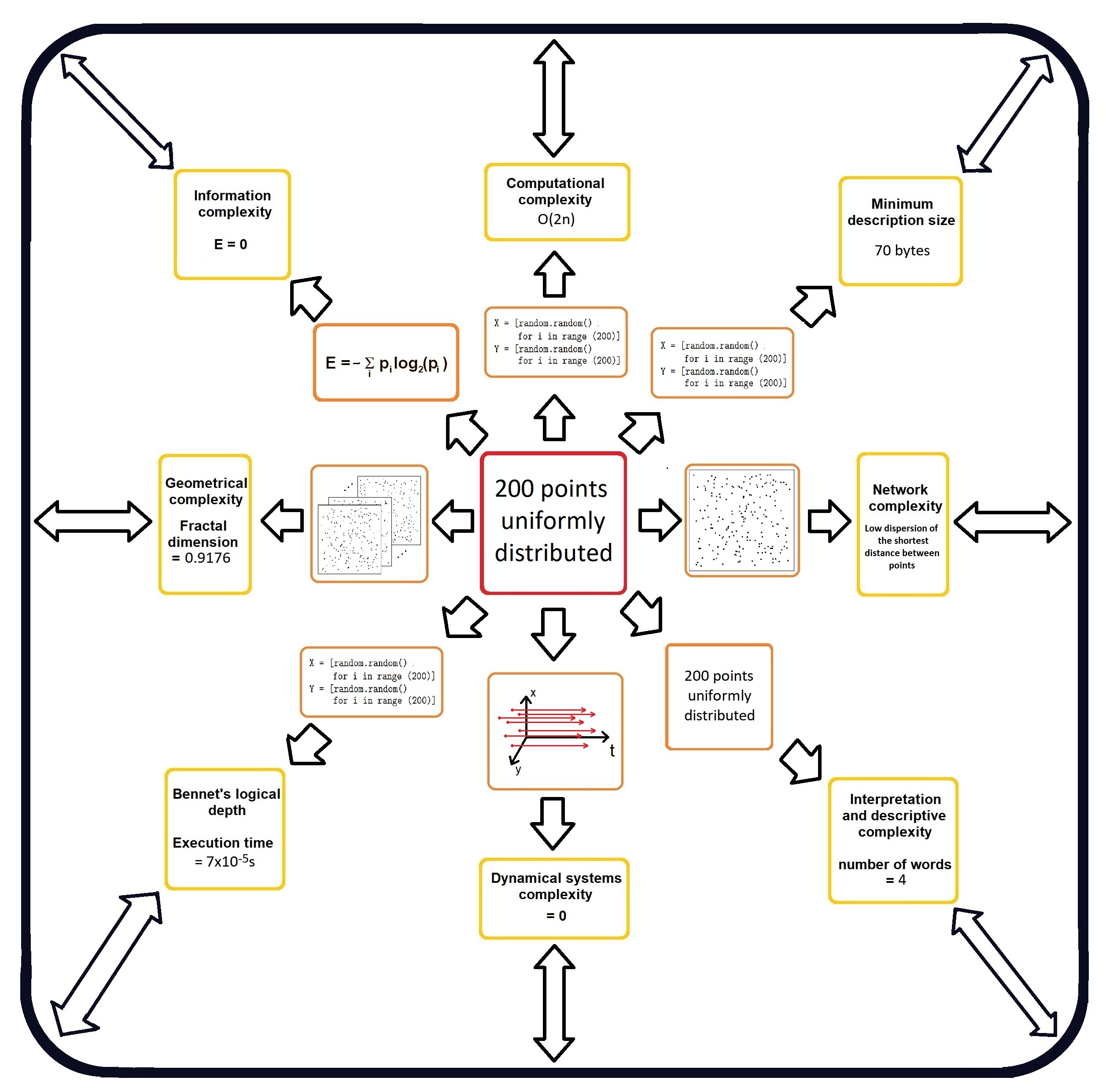}
		\caption{A possible overall framework integrating some of the several approaches to quantifying complexity, instantiated to the specific example in Figure~\ref{UniformDistribution}.}
		\label{Examples}
	\end{figure*}

	In this section, we briefly reviewed some of the several approaches at defining and quantifying complexity. Figure~\ref{Examples} illustrates the application of some of those complexity quantification methods.  A similar construction could be obtained for several other types of problems.  The enclosing interconnection pathway indicates that it is possible to interrelate and integrate the results obtained by different methods.  It remains a substantial challenge is to accommodate the discrepant indications of complexity as provided by these various approaches.
	
	\section{A cost approach to complexity}
	\label{Complexity}
	
	In the remainder of the present work, we aim at describing an approach to complexity that circumvents the problem that several of the other existing approaches of not being general enough, in the sense that counterexamples can often be found.  In other words, one of the main limitations of the previous approaches is that it is difficult to achieve a consensus, with a same problem being characterized as complex by one approach and as simple by another.  This partly follows from the fact that the several discussed approaches were developed having specific types of problems, systems and structures in mind.  
	
	In addition, a more general definition of complexity should be as much as possible compatible with its understanding by humans, which is intrinsically relative and can vary along time and space as well as in terms of resources availability.  So, a more comprehensive definition of complexity would need to rely on some quantification that can itself adapt given conditions varying in time and space.  As it happens, there is one concept that has been developed in order to account precisely for these characteristics, and which seems to provide an interesting approach to complexity.  We are referring to the concept of \emph{cost} as understood in economical sciences.  Indeed, cost is intrinsically adaptive to the availability of resources, reflecting also the most relevant constraints.  
	
	All in all, the main aspects of  the  proposed  approach include:  (a)  relating complexity to scientific modeling, in the sense that the  given  entity  whose  complexity  is  to  be  measured  is mapped  from  its specific  domain  into  a  respective  description (or model) in an abstraction (incorporating the mapping aspects from the Interpretation and Descriptive Complexity); (b) considering the non-bijective nature often characterizing such mappings, which implies in difficulties  to  recover/predict  the  original  entity (a  problem often studied in pattern recognition and computer vision, e.g.~\cite{abidi2016optimization});  (c) representing  both  the  original  object  and its respective  description  in  terms  of graphs/networks; (d) associating costs (e.g.  computational, economical or required for developing the model) to both the mapping and the errors incurred in recovering the original entity from its description or operating the resulting solution; and (e) probably more importantly, quantifying the complexity of the network representing the original entity in terms of costs.  
	
\begin{figure}[!bhtp]
		\centering
		\includegraphics[width=0.8\linewidth]{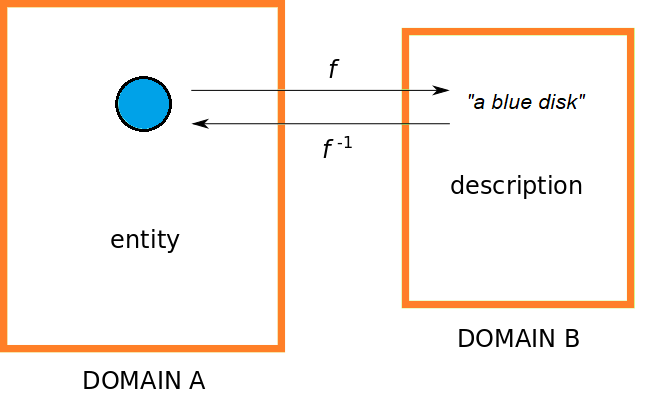}
		\caption{The modeling of an entity understood as a mapping $f$ of an entity from a  domain  A  (e.g. nature)  into  a respective  abstract description  in  domain  B (e.g.  English language).  In case the mapping is one-to-one (bijective), the original entity can be uniquely recovered through the respective inverse mapping $f^{-1}$. This is unlikely to occur in the real world, because there is a virtually infinite number of possible blue disks, so that the inverse mapping will be one-to-many and, therefore, non-bijective and non-invertible.}
		\label{bluedisk}
	\end{figure}

	Figure~\ref{bluedisk} illustrates the above aspect (a). Here, we have an entity in its original Domain A mapped by an application $f$ into a respective abstraction in Domain B.  For  instance,  Domain  A  could be the physical world,  while  Domain B would represent the respective description obtained by set of logic/mathematical/computational modeling approaches to be considered. Observe that, usually, the Domain B is more restricted than the Domain A, in the sense of containing fewer elements (simplifications), inherently implying the models to be incomplete. In  the  present  example,  a  blue  disk  is mapped into its linguistic description.  In case the inverse mapping $f^{-1}$ exists, it can be used to recover the original entity without any error.  However, this will not happen if Domain A is the real world, as there are virtually infinite possibilities of blue disks (e.g. varying in material or even slightly in color and texture, and/or presenting different sizes, not to mention the infinite range of details as one approaches the more microscopic scale).
	
	The above conceptualization is particularly helpful because it highlights the importance of the error in recovering of the original entity, which suggests that complexity would be related not only to developing a proper mapping $f$ and its inverse, but also  depend on the recovery error. In other words, larger reconstruction errors can be understood as indicatives not of the difficulty/complexity of modeling the original entity, but also of the difficulty of applying and operating the model.
	
	Indeed, the cost implied by modeling errors can be complemented by the cost of the \emph{operation} of the respectively developed solution, therefore also including other incurred expenses, such as maintenance, energy requirements, etc.  
	
	We  have  so  far  considered the object in Domain  A to be composed of a simple object as a circle. However,  most  of  real-world  objects  can  be understood  as  sets  of  components  interconnected  by  some relationships.   By  using  resources  such  as  semantic  networks  and  Petri  nets,  it  is  even  possible to  represent actions, procedures and programs as networks (e.g.~\cite{Arruda2019,peterson1981}).  
	
\begin{figure}[!htbp]
		\centering
		\includegraphics[width=0.8\linewidth]{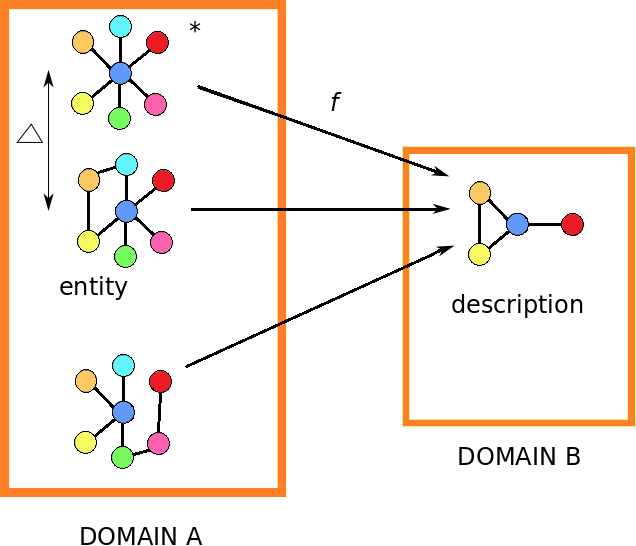}
		\caption{Modeling an entity represented by a respective network into a more abstract description involving a network.  Observe that the illustrated description is incomplete and not fully accurate, implying in the mapping $f$ being non-bijective.  Consequently, more than one entity in the Domain A can be mapped into the same representation in the Domain B, implying a degeneration in the mapping.  By imposing some additional restriction (e.g. regularization), it is possible to obtain a single inverse reconstruction (in this case, identified by the asterisk), whose error can be gauged in terms of some distance between the reconstructed and original entities.  In case $f$ is bijective, the mapping of the entity can be understood as being complete and necessarily invertible.  The higher the mapping cost $f$ and the error $\Delta$, the more complex the original entity would be.}
		\label{graphs}
	\end{figure}

	Figure~\ref{graphs} illustrates another example of modeling an entity, but now both the original object and  its  description  are  represented  as \emph{graphs/networks} (as in item (c) above). An immediate advantage of this approach is that some of the reasons for $f$ being non-bijective become evident:  the potential  complexity  of  the  entities  are  reflected in  the intricacy  of  the  respective  graphs.   In  addition,  entities having similar network representations (e.g.~differing by some missing connections or nodes), can be mapped into the  same  description  when $f$ fails  to  take  into  account such differences.  In the case of Figure~\ref{bluedisk}, this is reflected by the non-injective mapping of the three instances of the considered entity  into  the  same  representation.
	
	Directly  related  to the above discussed multiple mapping is the fact that the network respective to the description in the Domain B is simplified in the sense of being less complete than  the  network  representing  the  original  entity -- e.g.~by  having  nodes  with  properties different from that of the original entity  (colors  in  the case of the example in this figure) and/or missing connections or nodes.  
	
	There are, however, other possible sources of imprecision in the mapping, such as those implied by incorrect assumptions in the model construction, or also the presence of noise and incompleteness in the observations of the properties of the original entity.  This can also imply in a less accurate and less complete description being obtained in Domain B.	
	
	For all the reasons discussed above, the mapping $f$ can be  imprecise  and  non-bijective,  leading  to  errors  in  the reconstruction of the original entity from its description. In the case of being non-bijective, the inverse mapping can result in  more than one potential entity in Domain A, so that it is necessary to impose some restriction on the modeling (an approach knows as regularization, e.g.~\cite{Lu2013}), so that one of the recovered instances can be selected as being, potentially, the most likely and accurate. 
	
	Possible such restrictions  may  include  the  expected  number  of nodes, edges,  and/or  other  properties.   In  the  case  of  the  example in Figure~\ref{graphs}, the chosen inverse mapping, selected by the set of restrictions $R$,  is identified by an asterisk. The error $\Delta$ of the reconstruction can then be quantified by taking some distance between the original entity and the  selected  reconstruction.   It  seems  reasonable  to  understand that more complex entities will lead to less accurate mappings and descriptions, ultimately implying in larger reconstruction errors $\Delta$.  This line of reasoning leads to a possible alternative definition of complexity as:
	
	\begin{equation}
		\emph{complexity} \propto \left(\emph{cost}(f) + \emph{cost}(\Delta)\right)
	\end{equation}
	
	where $cost(f)$ is the development cost, and $cost(\Delta)$ is the error and operation expenses. In other words,  the complexity of an entity would be related  (not  necessarily  in  the  linear  sense)  to  the  sum of  the  cost  of  obtaining  the model,  as  well  as  the  cost  associated  to  the  error in the recovery (or prediction) of the original entity and operation of the respectively obtained solution.  We can also consider the following more general definition:
	
	\begin{equation}
		\emph{complexity} = g\left(\emph{cost}(f), \emph{cost}(\Delta)\right)
	\end{equation}
	
	where $g()$ is a potentially generic function integrating the modeling and operation costs.	
	
	The formula above reflects the hypothesis that  the  complexity  of  the  original  entity  or phenomenon would be given by a function $g()$ of the two considered costs, and very likely in such a way that higher development and operation costs would reflect in higher complexity.  This function is not fixed in order to better adapt to the demands and characteristics of each specific situation.
	
	Alternative operations could be considered to compose $g()$ in ways other than the sum of costs. However, the typically expected proportionality of the  complexity with both costs immediately discard some possibilities. Fractions are discarded by the fact that the complexity must increase with the costs (modeling and operational), having no inverse relationship with any type of cost. The product may also not be suitable, as even a zero operating cost, for instance, would imply in some complexity, as a perfect model (zero operating cost meaning that the model makes predictions with zero errors) must have a high complexity to be created, perhaps infinite.  An interesting alternative is the possibility to adopt a linear combination of costs, i.e.~$\alpha \emph{cost}(f) + \beta \emph{cost}(\Delta)$.
	
	An immediate advantage of the approach above is that it directly accommodates the often observed trade-off between these two costs, in the sense that more efforts are invested into developing a more complete and accurate model, therefore increasing $cost(f)$, the error and associated cost $cost(\Delta)$ tend to decrease.  On the contrary, in case the model is developed more quickly and less systematically, a larger error and operation cost will probably follow.   So, there seems to be a kind of trade-off between the costs $cost(f)$ and $cost(\Delta)$. This definition of cost for complexity takes into account the specificity of a model when applied to a given problem, and that cost may vary according to the problem complexity.  For instance, the most one learns about a given problem, the simpler it tends to become.
	
\begin{figure}[!tbp]
		\centering
		\includegraphics[width=0.8\linewidth]{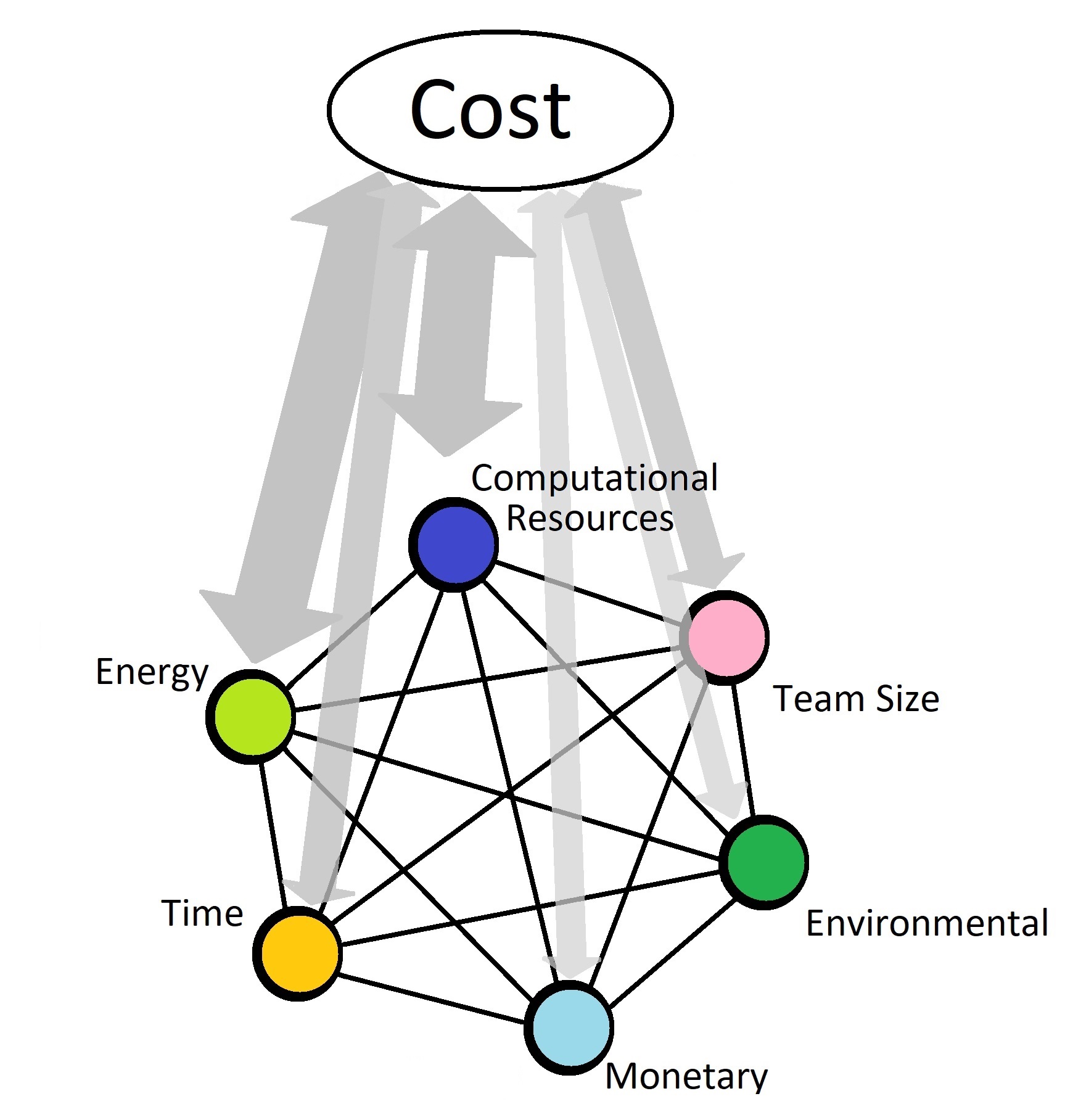}
		\caption{The modeling cost of an entity can be related to a wide range of possible effects, e.g. as an economic cost or as an energy cost, being more or less dependent of a chosen type. Different types of costs also can present an interdependence, varying with the application.}
		\label{Costs}
	\end{figure}

	Observe that the two involved costs can be defined in terms  of  several  aspects,  reflecting  each  specific  modeling problem.  For instance, we can take into account, as costs, the time (computational or taken for development) required for observing/measuring the original entity, obtaining/implementing $f$,  obtaining its  inverses,  and calculating the errors.   Alternatively, we could consider the computational complexity or the economical expenses required  for  the  modeling  project  (e.g.  wages,  resources, energy,  etc.). Costs of different natures are illustrated in Figure~\ref{Costs} and a  combination  of  these  costs  can  also  be adopted.  Interestingly, the choice of costs, and the costs themselves, can vary in time and space, but it is important that the reflect the specific demands and expectation while modeling or solving a specific problem.  For example, numerical computation was much more expensive and considerably  less  powerful  in  the  50’s  or 60’s  than  it  is  today.  In other words, what was complex in the past is likely to have  become  simpler.   
	
	Regarding  the  cost  to  be  associated  with  the  error/operation cost,  it  seems  to  be  reasonable  to understand that it is related (not necessarily in the linear way) to the reconstruction error $\Delta$, i.e.:
	
	\begin{equation}
		\emph{cost}(\Delta) \propto \Delta
	\end{equation}
	
	In particular, it is expected that the cost is zero when $\Delta$ is zero, which would seem to imply that the model is complete and therefore optimal given the design objectives.

	An  intrinsic  feature  of  the  here suggested approach to quantifying complexity is that it is potentially more closely related to the conceptual way in which us, humans, intuitively  tend  to  discern  between  complex  and  simple (at least in a more informal way).  In other words, when we say that a given entity, task or phenomenon is complex, we are inherently considering the expenses and/or difficulty required for its understanding (e.g. through studying and  modeling)  instead  of  some  more  abstract quantification such as derived from entropy or description length (though these aspects could also be complementary relevant).  In fact, probably we humans also consider these concepts by taking into account our previous modeling experiences with similar problems.
	
	The observed discrepancy between usual definitions of complexity and the human concept of it, as well as the difficult for quantify the costs involved, seems to indicate why there is an ambiguity and lack of consensus on the subject. As already suggested, complexity should not be understood as absolute mathematical quantification associated to a given model or problem, but remain relative in time and space, as in informal human conceptualization, allowing a given problem to be understood as more or less complex while taking into account the available techniques and instruments.  The proposed cost approach not only contemplates that demand, but can also be adapted to each situation, translating the simpler problem of how to quantify the involved costs.
	
	Another point to be taken into account is that a given problem often has many different solutions with respective advantages and disadvantages that may be more or less adequate to certain situations and, therefore, the complexity of a problem can also depend of the conditions that it are presented. For instance, when implementing a computational solution a common question is to choose either if that solution should prioritize a smaller execution time or a smaller use of memory or other resource. There is not an universal solution, but a set of solutions more adequate to specific demands of each situation.
	
	More complete examples of the application of the suggested cost approach to the characterization of the complexity, including real-world data, will be discussed in Section~\ref{sec:appls}.

	\section{Efficiency}
	
	Often, we enhance the complexity of models in the hope of obtaining a more complete representation, therefore diminishing the prediction error.  However, this increases the modeling cost. Therefore, the concept of \emph{efficiency} becomes important in order to quantify the advantage of more accurate modeling, or how much is gained by decreasing the error of the predictions of the model, which can be defined as a \emph{benefit}, at the expense of higher modeling costs. 
	
	The concept of \emph{benefit} is intrinsically relative to each situation, and needs to be specified respectively.  Examples of benefit would be how much the quality of living may increase by a new product, or how much a new bridge may contribute to reducing the average traveling time, the increase of accuracy, or the reduction of the chances of operation errors.  Often what is sought is a combination of these improvements.
	
	A possible approach to quantifying efficiency of the modeling approach could be
	
	\begin{equation}
		\emph{efficiency} \propto \frac{\emph{benefit}}{\emph{cost}}
	\end{equation}
	
	where \emph{cost} refers to the overall cost of the solution, including modeling and operation.
	
	Considering that the overall cost is related to the complexity of the solution (as discussed in Section~\ref{Complexity}), we can rewrite the previous definition of efficiency as
	
	\begin{equation}
		\emph{efficiency} \propto \frac{\emph{benefit}}{\emph{complexity}}
	\end{equation}

	\section{Case-examples of the cost approach} \label{sec:appls}
	
	\subsection{Statistic distribution}
	
	Given a generic function $P(x)$  sampled at a sufficient number of values of $x$ and respective ordinates $y = P(x)$, forming a set of samples represented in terms of \emph{Dirac's delta functions $\delta(x)$}, it is often necessary to obtain an interpolation between these sampled points.  This can be done in several manners, including by applying a convolution~\cite{Costa2019convolution} between that set of $\delta$ and a kernel such as the \emph{Gaussian function}, so as to interpolate smoothly between the missing information (gaps).
	This provides a simple example of the concept of regularization, in which smoothness is adopted as a constraint to complement the missing information in a dataset.
	The convolution between a Gaussian $G(X)$ with standard correlation matrix $K$ and a set of $N$ Dirac's delta functions $\delta(X-x_i)$ with the set of positions $X = [x_1, x_2, ... ,x_N]$, denoted as $(\delta*G)(X)$, consists of a sum of element-wise products of the set of $\delta(X-x_i)$ and $G(X)$ displaced by $j$ over each $\delta$ of the set. That is
	
	\begin{equation}
		(\delta*G)(X) = \sum_{i=1}^{N}{G(X - x_i)}
	\end{equation}
	
	It should be observed that the above equation refers to a simplification of the more general convolution expression, allowed by the symmetry of the Gaussian function and the sampling property of the Dirac's delta function (e.g.~\cite{Costa2019convolution}).
	
	\begin{figure*}[!tbp]
		\centering
		\includegraphics[width=0.7\linewidth]{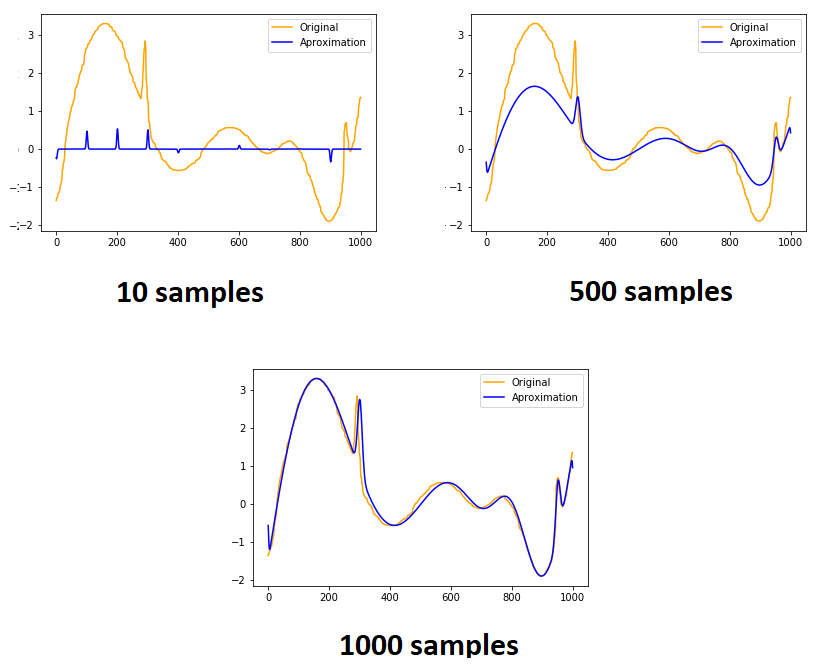}
		\caption{The convolution of a set of \emph{Dirac's delta functions} and a \emph{Gaussian function} approximate of the original function $P(x)$ with a precision that depends of the number of samples taken.}
		\label{fdp}
	\end{figure*}
	
	The above principle is also called a \emph{Kernel density estimation}~\cite{parzen1962} and its outcome consists of a new function that approximates $P(x)$ with a precision that depends on the number of samples taken, as can be seen in Figure~\ref{fdp}.
	
	\begin{figure}[!thbp]
		\centering
		\includegraphics[width=0.9\linewidth]{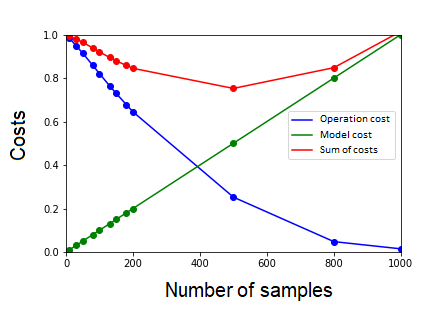}
		\caption{The relationship of modeling and operation costs, as well as their sum, for different numbers of samples taken for reconstruct a statistic distribution by convolution. The costs have been normalized so as to be within the interval $[0,1]$}.
		\label{fdp_cost}
	\end{figure}
	
	We can understand the computational cost (processing and storage) as our modelling cost, depending of the number of samples taken, while the operation cost of our model could correspond to the quadratic error between the convolution outcome and $P(x)$. By normalizing the modelling and operation costs for different numbers of samples, as shown in Figure~\ref{fdp_cost}, we can analyse the relationship between the involved costs.
	
	As this result shows, the model cost increases with the number of samples, as expected, reflecting the increasing complexity of the model. However, the error decreases with the number of samples, implying the total cost to reach a minimum for $500$ samples, indicating an optimal sample size (for this example) to reduce the error of approximation on the face of the cost of applying a higher number of samples, as beyond the mark of $500$ samples the efficiency of the solutions tends to decrease, producing a more complex model without effective gain in accuracy, considering the adopted hypothesis.

	\subsection{Simulated annealing}
	
	Inspired on metallurgic processes, simulated annealing~\cite{press1992,kirkpatrick1984} consists of an optimization method that incorporates a temperature-like parameter and relative variations of an objective function. The objective function is a measurement of the effectiveness of a solution of a problem, being often understood in terms of an associated energy. The above mentioned temperature parameter $T$ is used to control the probability of making a modification leading to higher energy, and is progressively decreased according to a certain strategy. The decision to take specific configurations is based on the Boltzmann distribution:
	
	\begin{equation}
		p_i = Z \exp(- \frac{\Delta E_i}{kT})
	\end{equation}
	
	where $Z$ is a normalization constant, $\Delta E_i$ is the objective function variation, and $k$ is the Boltzmann constant.  The progressive reduction of $T$ implies smaller probabilities of taking choices leading to higher energy, which are necessary for eventual convergence to the global extreme.
	
	Simulated annealing can be used to improve the \emph{gradient descent method}~\cite{press1992} by allowing larger and less direct steps to be taken at high $T$ to avoid local minima.  The gradient descent can be associated to the trajectory of an agent that moves according to the gradient of the given field.   As $T$ decreases, the descent becomes more controlled and similar to the traditional gradient descent, therefore increasing the chances of reaching a suitable minima.
	
	\begin{figure}[!htbp]
		\centering
		\includegraphics[width=0.9\linewidth]{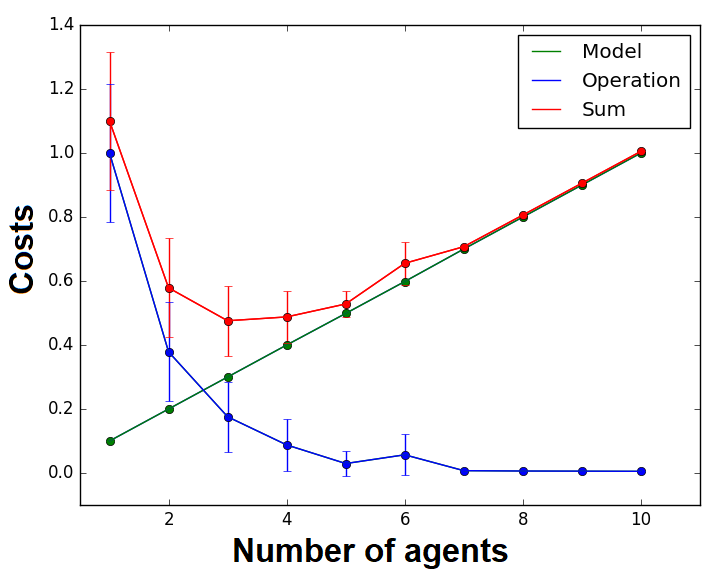}
		\caption{The modeling and operation costs in terms of the number of employed agents, as well as the sum of these costs, obtained while trying to find, by using simulated annealing gradient descent, the overall minimum of a scalar field corresponding to a linear combination of Gaussian functions. The costs have been normalized so as to be within the interval $[0,1]$ and the vertical error bars are depicted to only $20\%$ of its real size.}
		\label{Annealing}
	\end{figure}
	
	As our second cost approach example, we consider a surface generated by a linear combination of five three-dimensional Gaussian functions, with the same covariance matrices and different magnitudes and centers, and perform gradient descent with simulated annealing in order to try to find the global minimum value.   The temperature decrease strategy consisted of subtracting $10\%$ from the current temperature after each $100$ steps of simulation. 
	
	As modeling cost, we take the sum of times spent by each agent while searching for the minimum, and for operation cost the Euclidean distance between the known global minimum point and the closest answer of the simulated agents. The results can be seen in Figure~\ref{Annealing} in terms of the number of agents employed to simultaneously perform the simulated annealing-controlled gradient descent.
	
	The total cost reaches its minimum value for  $3$ agents (or $4$ agents if we consider probabilistic fluctuations proportional to the error bar), suggesting that an increase in complexity (implied by operation with more agents) would lead to lesser efficiency.

	\subsection{Airport network}
	
	In order to study a real-world example, we considered an airport network that represents flight connections in the United States.~\cite{transtatsNetwork} Considering the variation of the airplane fuel-per-gallon price between years $2000$ and $2019$~\cite{transtatsData} as the operation cost, we can vary the number of edges (model cost - calculated as the total length of the edges of the network) starting from the minimal spanning tree~\cite{graham1985} of the network, using the distances between airports as respective edge weights. 
	
\begin{figure}[!htbp]
		\centering
		\includegraphics[width=\linewidth]{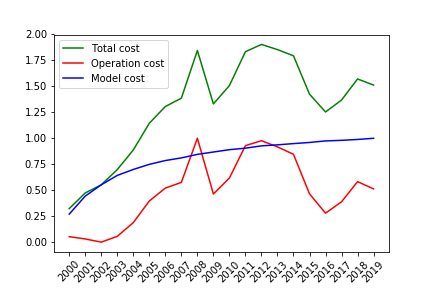}
		\caption{The modeling and operation costs, as well as their sum, obtained for the total length of edges and the fuel price, respectively, when considering modifications on the minimum spanning tree of the airport network and the variation of the fuel price. The minimum value of the model cost corresponds to the minimum spanning tree total length, while the maximum value stand for the original network total length.  The operation and model costs have been normalized so as to remain within the interval $[0,1]$}.
		\label{airports}
	\end{figure}

	For an experiment with the airport network we can reintroduce portions of the original edges into the minimal spanning tree, as to represent an improvement to the topology of the network across the years, at the expense of additional model cost. A fraction of the original edges is randomly chosen and added into the network at each time step, yielding the interrelation between the modeling and operation costs as can be seen in Figure~\ref{airports}. For a reduced topology satisfying the problem and a reduced fuel cost, as in the left end of the curve, this system is perceived as simpler than when a higher number of airlines is needed to satisfy a solution or when the costs of operate the system increases. The need of more airlines, to deal with an increasing number of passengers or to better administrate longer flights, is a reflex of a more complex problem and, therefore, has greater costs involved.

	\subsection{Kuramoto model}
	
	The Kuramoto model~\cite{rodrigues2016} is a dynamic model that reproduce the appearance of synchronization effects on systems of many elements having some coupling. Typically an example of application of this model is the set of $N$ coupled harmonic oscillators, represented with phases $\theta_i(t)$ and frequencies $\omega_i$ for each oscillator. By calculating the governing equations of this set
	
	\begin{equation} \label{Kuramoto}
		\frac{d\theta_i}{dt} = \omega_i + R\lambda sin(\psi - \theta_i)
	\end{equation}
	
	where $\psi(t)$ is the average phase of the oscillators and $\lambda$ is the coupling force, we can estimate the time required for system synchronization, in case it exists, as measured by the order parameter $R$, given as
	
	\begin{equation}
		Re^{i\psi(t)} = \frac{1}{N} \sum_{j=0}^N e^{i\theta_j(t)}
	\end{equation}
	
\begin{figure}[!htbp]
		\centering
		\includegraphics[width=0.9\linewidth]{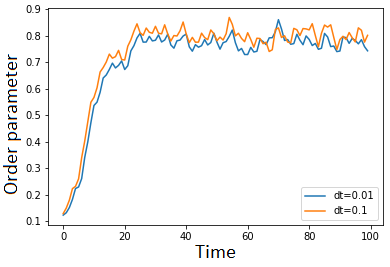}
		\caption{Evolution of the order parameter $R$ of the Kuramoto model for $100$ oscillators, calculated for two different time steps $dt$. These curves correspond to the mean values of $R$ for $1000$ repetitions each.}
		\label{parametroOrdem}
	\end{figure}

\begin{figure}[!htbp]
		\centering
		\includegraphics[width=0.9\linewidth]{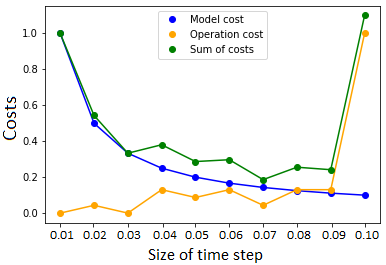}
		\caption{The modeling and operation costs, as well as their sum, obtained for the Kuramoto model for $100$ oscillators, calculated for different time steps $dt$ with the Heun's method. The costs have been normalized so as to remain within the interval $[0,1]$.}
		\label{KuraResultados}
	\end{figure}

	We can calculate Equation~\ref{Kuramoto} using the Heun differential equation numerical method,~\cite{butcher2008} for instance, varying the size of increment in time $dt$ in different realizations and adopt the computational cost as our modelling cost, while the operation cost could correspond to the difference between the time necessary for the order parameter to reach $0.8$ relative to the case of $dt=0.01$, which is assumed to be the more precise among the considered cases. The mean values of the evolution of the order parameter $R$ for $1000$ repetitions of $100$ oscillators, calculated for two representative time steps can be seen in Figure~\ref{parametroOrdem}, while the normalized modelling, operation and total costs are shown in Figure~\ref{KuraResultados}.  Observe that the error (operation cost) increases as the model cost diminishes, presenting a sharp divergence near step size 0.09  while the total cost, or complexity, of this modeling has two major peaks, representing the situations where this solution for the problem is considered complex as it is not enough for reaching accurate predictions (the operation cost is too high in the case of the present approach) or where too much computational power was employed to solve it, making it a complex task (high model cost).
	
	\section{Concluding remarks}
	
	Complexity is a word that has been recurrent and frequently used not only in science, but also in virtually all human-related activities.  Yet, it remains an elusive concept, despite a large number of respective approaches, to obtain a respective definition flexible enough to account to most of the structures and dynamics normally considered complex.
	
	In the present work, we have revised how complexity has been approached from several points of view, from entropy to description length. That  a  more  complete  understanding  of  complexity  involves  so  many  aspects  is  hardly  surprising,  given  that complexity is inherently subjective and challenging. So, we have discussed complexity considered from  perspectives including data and coding size/length, geometrical intricacy, critical divergence of  dynamics,  and  network  topology.   Tough each  of  these  approaches offers its intrinsic contribution to better understanding  and  quantifying  complexity  while  studying  an entity and/or dynamics, they tend to be specific to the type of problems and area in which they were developed.  
	
	In addition to reviewing some of the many insightful ways in which complexity has been characterized, we also tried to integrate several of the principles underlying  these  approaches,  as  well  as  incorporate  concepts from areas such as pattern recognition and network science,  into a  more  conceptual and general  model  of  complexity  which is primarily based on the completeness of representations understood as mapping of an entity from a domain into another.  In addition,  concepts from scientific modeling, pattern  recognition  and  network  science  were  also  incorporated, giving rise to an approach in which the complexity of an entity can be understood in terms of the cost of obtaining a proper mapping and the cost implied by the almost  unavoidable  reconstruction  errors and operation. The adoption of cost is particularly interesting because this concept has been developed along ages in economy-related areas precisely as a quantification of the difficulty or scarcity of resources that can adapt along time and space, while also reflecting distinct specific demands.
	
	After presenting and discussing the cost-based approach to complexity, we also provided a sequence of examples ranging from the simple random field example presented at the beginning of this work to situations involving algorithms and real-world data.  In all these considered situations, the proposed cost-based approach provided an effective, flexible, and general conceptualization and quantification of the respectively involved complexities.

	\acknowledgements    
	L. da F. Costa thanks CNPq (307333/2013-2) and FAPESP (2011/50761-2) for support. G. S. Domingues thanks CNPQ (131909/2019-3) for financial support. This study was financed in part by the Coordenação de Aperfeiçoamento de Pessoal de Nível Superior – Brasil (CAPES) – Finance Code 001.

\newpage
\nocite{*}

\end{document}